\documentclass[aip,apl,10pt,twocolumn,floatfix]{revtex4-1}

\usepackage{amsmath}
\usepackage{graphics}

\include{MyMc}

\newcommand{\Sec}[1]{}
\newcommand{\Subsec}[1]{}

\begin{document}

\title{%
  Radiation power and linewidth of a semifluxon-based Josephson oscillator
}

\author{M. Paramonov}
\author{M. Yu. Fominsky}
\author{V. P. Koshelets}
\affiliation{%
  Kotel'nikov Institute of Radioengineering and Electronics RAS, 
  Mokhovaya 11, 125009 Moscow, Russia
}

\author{B. Neumeier}
\author{D. Koelle}
\author{R. Kleiner}
\affiliation{%
  Physikalisches Institut and Center for Collective Quantum Phenomena in LISA$^+$,
  Universit\"at T\"ubingen, Auf der Morgenstelle 14, D-72076 T\"ubingen, Germany
}

\author{E. Goldobin}
\altaffiliation[Current address:]{
  Center of Interdisciplinary Studies, 
  Ostmarkstrasse 8, 72135 Dettenhausen, Germany%
}
\affiliation{%
  Physikalisches Institut and Center for Collective Quantum Phenomena in LISA$^+$,
  Universit\"at T\"ubingen, Auf der Morgenstelle 14, D-72076 T\"ubingen, Germany
}

\date{%
  \today\ File: \textbf{\jobname.\TeX}
}

\begin{abstract}
  We demonstrate a high-frequency generator operating at $\sim 200\units{GHz}$ based on flipping a semifluxon in a Josephson junction of moderate normalized length. The semifluxon spontaneously appears at the $\pi$ discontinuity of the Josephson phase artificially created by means of two tiny current injectors. The radiation is detected by an on-chip detector (tunnel junction). The estimated radiation power (at the detector) is $\sim 8 \units{nW}$ and should be compared with the dc power of $\sim 100\units{nW}$ consumed by the generator. The measured radiation linewidth, as low as $1.1\units{MHz}$, is typical for geometrical (Fiske) resonances although we tried to suppress such resonances by placing well-matched microwave transformers at its both ends. Making use of a phase-locking feedback loop we are able to reduce the radiation linewidth to less than $1\units{Hz}$ measured relative to the reference oscillator and defined just by the resolution of our measurement setup. 
\end{abstract}



\maketitle

\Sec{Introduction}
\label{Sec:Intro}

Josephson oscillators based on the shuttling of fluxons along a long Josephson junction (LJJ) in the zero-field step (ZFS) mode were investigated in the 1980s\cite{Dueholm:1981:LJJ:Multisolitons,Joergensen:1982:ZFS-LW,Soerensen:1984:ZFS/FS:Radiation}. Unfortunately such oscillators cannot deliver appreciable power to the load because of contradictory requirements. On one hand, to maximize the power delivered from a LJJ to the output microwave line the coupling between the output line and the LJJ should be strong (impedance matching). On the other hand, for such a good coupling, the fluxon will not reflect from the edge of the LJJ. It will annihilate (i.e. be lost), thus stopping the generation.

Currently, Josephson oscillators are mainly used in the flux-flow regime\cite{Nagatsuma:1983:FFO-I,Nagatsuma:1984:FFO-II,Nagatsuma:1985:FFO-III,Koshelets:2000:IntRecv,Koshelets:2007:IntRec4TELIS} optionally at the Fiske (geometrical) resonances. In this regime the power delivered to the load can reach $\sim 1\units{\mu W}$, an efficiency of dc to ac conversion $\sim10\units{\%}$, and a free-running radiation linewidth $\sim 1\units{MHz}$. The linewidth depends on the differential resistance at the working point and is sensitive to the bias-current noise as well as to the field (control line current) noise.

Nowadays one can create artificial phase discontinuities in LJJs\cite{Ustinov:2002:ALJJ:InsFluxon,Goldobin:2004:Art-0-pi}. After introducing a $\pi$ discontinuity of the Josephson phase, the LJJ reacts by creating a semifluxon\cite{Xu:SF-shape,Goldobin:SF-Shape,Hilgenkamp:zigzag:SF} pinned at such a discontinuity. A semifluxon being biased over its depinning current\cite{Susanto:SF-gamma_c} ($=(2/\pi)I_c$ in an infinite LJJ, where $I_c=j_c A$ is the ``intrinsic'' critical current, $j_c$ is the critical current density, $A$ is the area of the LJJ) flips continuously between the semifluxon and antisemifluxon states\cite{Goldobin:SF-ReArrange}. Upon each flip, an integer (anti)fluxon is emitted and moves under the action of the bias current towards the left (right) edge of the LJJ. Usually\footnote{In some non-Lorenz invariant systems the Swihart velocity can be exceeded and one observes a Cherenkov radiation tail behind the fluxon\cite{Goldobin:Cherry2}.} the bias current can accelerate the (anti)fluxons up to the Swihart velocity and a semifluxon voltage step (half-integer ZFS) appears on the current-voltage characteristic of the LJJ\cite{Stefanakis:ZFS/2}. Although such semi-integer ZFS were observed experimentally\cite{Goldobin:2004:Art-0-pi,Pfeiffer:2008:SIFS-0-pi:HIZFS} no measurements of the radiation power or the radiation linewidth were carried out. If the LJJ is biased to the half-integer ZFS, the end of the LJJ can be well coupled to the output line so that the arriving (anti)fluxon can even be absorbed and a considerable part of its energy is emitted into the output line. Still, the generation will not stop like in the case of ZFS described above. It will continue as it originates from the center of the LJJ. Thus, such a generator based on a flipping semifluxon should provide much better output power and dc-to-ac energy conversion efficiency.

In this letter we report on the high-frequency study of such a generator based on a flipping semifluxon. We couple it to a detector and investigate the power delivered to the detector, dc-to-ac power conversion efficiency as well as radiation linewidth of such a generator. Finally, we demonstrate that using a phase-locking feedback loop one can reduce the linewidth practically to zero, defined just be the accuracy of our measurement equipment ($\sim 1\units{Hz}$).

\Sec{Experiment}
\label{Sec:Exp}

\Subsec{Samples}

\begin{figure}[!tb]
  \centering\includegraphics{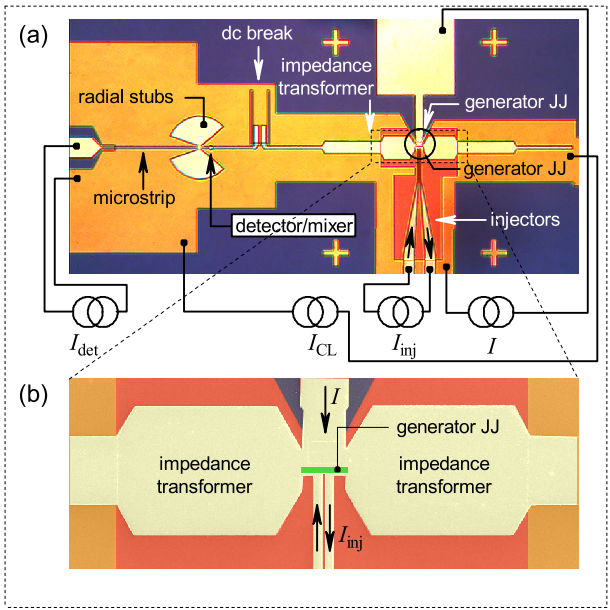}
  \caption{(Color online)
    (a) Optical photograph of the whole device \#10. (b) Zoomed picture of the generator JJ obtained by means of electronic microscope/lithograph \textsc{Raith} e-LiNE (colored manually). 
  }
  \label{Fig:Photo}
\end{figure}

We have fabricated several samples using conventional Nb$|$AlO$_x|$Nb JJ technology\cite{Koshelets:1991:Nb-tech,Dmitriev:2003:Nb-tech}. Each device consists of a generator JJ equipped by a pair of current injectors, see Fig.~\ref{Fig:Photo}. Injectors are connected to the center of the top superconducting electrode by microstrip lines that are equipped with high-frequency filters to avoid escape of the rf power along injector lines (filters are situated outside of the Fig.~\ref{Fig:Photo}(a) frame). The samples have different generator JJ lengths $L$, width $w$, as well as different injectors widths $\Dx$ and gaps $\Dx$ between them. The bottom electrode of the generator JJ is embedded into a control line connected in-line (along the length of the JJ) to produce a magnetic field. Each edge of the generator JJ is coupled to the high frequency coupling circuit. On one (dead) end it contains only an impedance transformer. On the other (active) end used for detection and linewidth measurements it contains two impedance transformers and a dc break. The coupling circuit is calculated to transmit the emitted radiation in a certain frequency range (typical bandwidth  $\sim 100\units{GHz}$ at a frequency range of $200$--$500\units{GHz}$). The active coupling circuit was finally connected to an on-chip detector/mixer equipped with a special tuning circuit. A tuning circuit consists of an inductance (a piece of a microstrip) and radial stubs. It is used to tune out the detector junction capacitance to provide a \emph{real} impedance at the frequency of the generator. The detector is a small $\sim 2\units{\mu m^2}$ tunnel JJ, which was connected to an external reference rf source (local oscillator) by means of a microstrip line. This microstrip is used (a) to dc bias the detector junction, (b) to supply an external reference (local oscillator) signal and (c) to send the IF signal from the detector/mixer to the external spectrum analyzer. The parameters of the samples discussed in this paper are summarized in Tab.~\ref{Tab:samples}. The particular fabrication run discussed here delivered $j_c\approx 1.7\units{kA/cm^2}$, which gives the Josephson length $\lambda_J\approx 10\units{\mu m}$. For all the samples discussed here the voltage of the semifluxon step (the same as the voltage of the first Fiske step) $V_\mathrm{FS1}\approx400\units{\mu V}$.

\begin{table}
  \begin{tabular}{lcccc}
    name & $L (\units{\mu m})$ & $w (\units{\mu m})$ & $\Delta w (\units{\mu m})$ & $\Delta x (\units{\mu m})$\\
    \hline
    \#02 & 15.8 & 1.4=2.0-0.6 & 2.5=2.0+0.5 & 1.5=2.0-0.5\\
    \#03 & 15.8 & 1.9=2.5-0.6 & 2.5=2.0+0.5 & 1.5=2.0-0.5\\
    \#10 & 15.8 & 2.4=3.0-0.6 & 2.5=2.0+0.5 & 1.5=2.0-0.5\\
    \hline
  \end{tabular}
  \caption{%
    Parameters of the devices mentioned in the text. Given sizes represent the target values that are the design values (in photomask file) plus technological corrections: the widths of the JJs comes $\approx0.6\units{\mu m}$ smaller than in the file, widths of the injector lines come $\sim 0.5\units{\mu m}$ larger than in the file).
  }
  \label{Tab:samples}
\end{table}

\Subsec{Experiment}

\begin{figure}[!tb]
  \centering\includegraphics{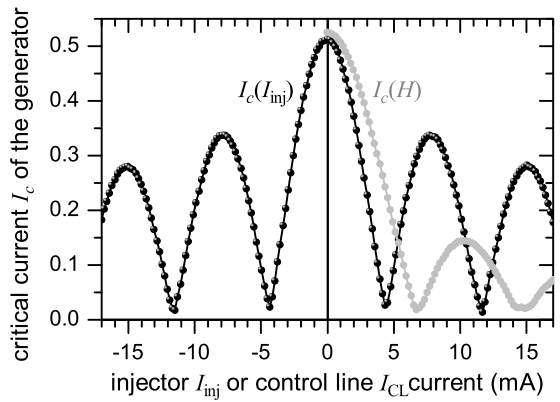}
  \caption{%
    The dependence $I_c(I_\mathrm{inj})$ (black) of the generator JJ used for injector calibration and determination of $I_\mathrm{inj}^\mathrm{\pm\pi}$. The dependence $I_c(I_\mathrm{CL})$ (gray) looks similar to a Fraunhofer pattern demonstrating good uniformity of the generator JJ. Both dependences are obtained for the sample \#03, see Tab.~\ref{Tab:samples}. 
  }
  \label{Fig:GenJJ:I_c(I_inj)}
\end{figure}

First, we have calibrated the injectors by measuring the dependence of the critical current of the generator JJ vs. injector current, \ie, $I_c(I_\mathrm{inj})$. As predicted\cite{Gaber:2005:NonIdealInj2}, it looks like an almost periodic function with parabolic maxima and cusp-like minima as shown in Fig.~\ref{Fig:GenJJ:I_c(I_inj)}. The ``period'' corresponds to a phase discontinuity $\kappa \propto I_\mathrm{inj}$ changed by $2\pi$. This allows to calibrate injectors, \ie, to determine the proportionality coefficient between $I_\mathrm{inj}$ and $\kappa$. In the case shown in Fig.~\ref{Fig:GenJJ:I_c(I_inj)}, the $\pi$ discontinuity needed to create a semifluxon is reached for $I_\mathrm{inj}=I_\mathrm{inj}^\mathrm{\pm\pi}\approx 4.4\units{mA}$. The decrease of the amplitude of $I_c(I_\mathrm{inj})$ at the $\pm1^\mathrm{st}$ maxima is related to the finite size of injectors\cite{Gaber:2005:NonIdealInj2}.

%

\begin{figure}[!tb]
  \centering\includegraphics{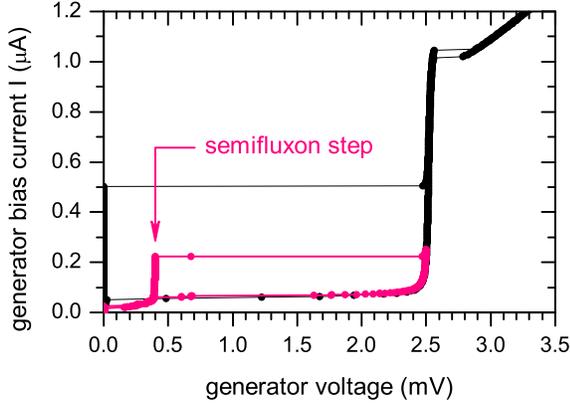}
  \caption{(Color online)
    IVCs of generator JJ with $I_\mathrm{inj}=0$ (black) and $I_\mathrm{inj}=I_\mathrm{inj}^\pi$ (pink/gray) from sample \#03. 
  }
  \label{Fig:GenJJ:IVC}
\end{figure}

Second, we have measured the $I$--$V$ characteristic (IVC) of the generator JJ at $I_\mathrm{inj}=I_\mathrm{inj}^\pi$ and compared it with the IVC measured at $I_\mathrm{inj}=0$, which looks like the usual IVC of a tunnel JJ, see Fig.~\ref{Fig:GenJJ:IVC}. At $I_\mathrm{inj}^\pi$ a $\pi$ discontinuity of the phase is produced. This results in the creation of a(n) (anti)semifluxon pinned at it. Actually, the localized flux is smaller than $\Phi_0/2$ because the normalized length of the JJ is a few $\lambda_J$ instead of $\infty$ ($\Phi_0\approx2.07\times10^{-15}\units{Wb}$ is a magnetic flux quantum). In this case the critical current of the JJ is, in fact, a depinning current of the (anti)semifluxon and is much lower than $I_c$ at $I_\mathrm{inj}=0$, see Fig.~\ref{Fig:GenJJ:IVC}. Upon exceeding the critical current the generator JJ jumps to the semifluxon step aka half-integer zero-field step (HiZFS)\cite{Stefanakis:ZFS/2,Goldobin:2004:Art-0-pi,Pfeiffer:2008:SIFS-0-pi:HIZFS}. It is the dynamics at this step that is the main subject of our study. 

\begin{figure}[!tb]
  \centering\includegraphics{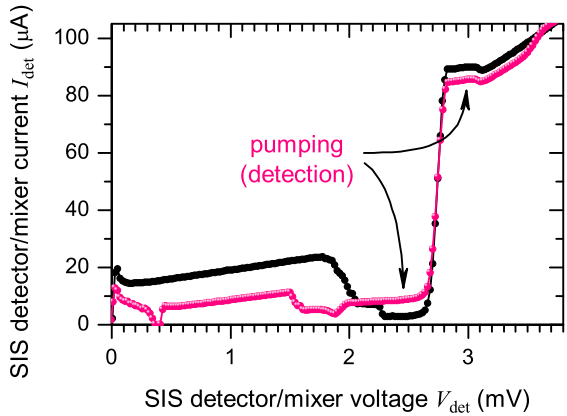}
  \caption{(Color online)
    Autonomous (black) and pumped (pink/gray) IVC of the detector JJ from sample \#02 measured in the voltage-bias mode. 
  }
  \label{Fig:DetJJ:IVC}
\end{figure}

Third, we choose a working point of the generator JJ somewhere at the semifluxon step and measure the IVC of the detector JJ. An example of such an IVC of a pumped detector is presented in Fig.~\ref{Fig:DetJJ:IVC} in comparison with the unpumped detector, \ie, when the generator JJ is at $I=V=0$. One can observe that in the vicinity of the gap voltage $V_g\approx 2.7\units{mV}$ the quasiparticle current increases (for detector voltage $V_\mathrm{det}<V_g$) or decreases (for $V_\mathrm{det}>V_g$). By fitting the IVC of the pumped detector using a Tien-Gordon model\cite{Tien-Gordon:1963:SIS+Radiation} we can estimate the ac power $P_\mathrm{det}^\mathrm{ac}$ delivered to the detector. In the typical case $P_\mathrm{det}^\mathrm{ac} \sim 8\units{nW}$. The dc power $P_\mathrm{gen}^\mathrm{dc}$ consumed by the generator at the working point situated at the semifluxon step is $V\times I \sim 400\units{\mu V}\times 200 \units{\mu A}\sim 80 \units{nW}$. These figures let us conclude that the generator makes a rather efficient conversion of dc power to ac power. At this point quasi-dc measurements were complete.

\begin{figure}[!tb]
  \centering\includegraphics{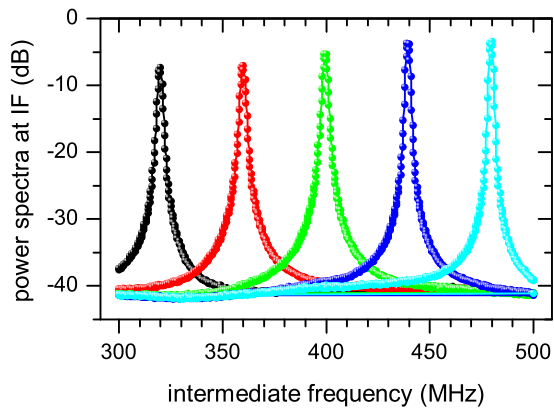}
  \caption{(Color online)
    Emission spectra measured at intermediate frequency at different bias points of the generator JJ (sample \#02 
    ) at a frequency of about $197\units{GHz}$. The mixer works at the $N=16$ harmonic of the local oscillator frequency  $f=12.283\units{GHz}$. The radiation linewidths $\Delta f$ of $2.0\pm 0.2\units{MHz}$ are measured at slightly different bias points $I$ in the range $338$--$345\units{\mu A}$.
  }
  \label{Fig:LW}
\end{figure}

Forth, to measure the emitted radiation linewidth the device (chip) was mounted in a specially designed high frequency cryostat described in detail elsewhere\cite{Koshelets:2000:PLL,Koshelets:2000:IntRecv}. The generator JJ was biased to the semifluxon step exactly as described above. The detector JJ was not dc biased (for $N$ even, see below) or somewhat biased (for odd $N$)\cite{Kalashnikov:2011:HarmonicMixer}. An external reference rf source (local oscillator) \textsc{Rhode\&Schwarz SMP03} sends a microwave power at a frequency $\sim 10$--$20\units{GHz}$ via microstrip line to the detector JJ, which now works as a high-harmonic mixer (non-linear element). The $N$-th harmonic ($N\sim 10$--$20$) of the local oscillator mixes with the signal arriving from the semifluxon generator. The resulting down-converted signal at the intermediate frequency (IF) is received by the room-temperature spectrum analyzer \textsc{HP E4440A}. Fig.~\ref{Fig:LW} shows several radiation spectra taken at closely situated bias points at the semifluxon step. We were able to continuously tune the central frequency of the radiation peak by changing the bias current of the generator JJ. The power $P_\mathrm{det}^\mathrm{ac}$ (height of the peak) and the radiation linewidth are weak functions of the generator bias current. Among all samples the values of linewidth $\Delta f$ between $1.1$ and $10\units{MHz}$ is observed over the full range of the bias current. Such values of the linewidth are typical for generators oscillating at geometrical (Fiske) resonances, as in our case the frequency of the semifluxon step nominally coincides with the frequency of the first Fiske mode. We note, however, that we tried to suppress somewhat the Fiske resonance by inserting well-matched microwave transformers at both ends of the generator JJ, see Fig.~\ref{Fig:Photo}. Upon deviations of $I_\mathrm{inj}$ from $I_\mathrm{inj}^\pi$ the radiation frequency and the linewidth do not change substantially until $I_\mathrm{inj}$ changes so much that the semifluxon step disappears (or the bias point jumps from it). This is in contrast to the generators working in the flow-flow mode. Their frequency depends on applied magnetic field $H\propto I_\mathrm{CL}$ and the linewidth is, therefore, sensitive to $I_\mathrm{CL}$ noise\cite{Koshelets:2001:FFO-linewidth}.  

Finally, a proven technique to reduce the radiation linewidth is to use a feedback phase-locking loop (PLL)\cite{Koshelets:2000:PLL}. To be effective the bandwidth of the PLL circuitry should exceed the linewidth of the free-running generator. The PLL's feedback signal was added to the bias current $I$ of the generator. After this the linewidth, measured relative to the reference oscillator, collapses almost to zero having the spectral ratio of 95\% (the ratio of the power in the main narrow peak to the total emitted power). The remaining linewidth is defined just by the accuracy of our instrumentation ($\sim 1\units{Hz}$). Further details of the operation of semifluxon generator with PLL will be presented elsewhere.

\Sec{Conclusions}

In summary, we have investigated the output power and the radiation linewidth of a high frequency cryogenic generator based on flipping semifluxon in a Josephson junction of moderate normalized length $L/\lambda_J\sim2$. The generator delivers an output power of $\sim 8\units{nW}$ measured at the detector. It also shows a good conversion efficiency $\sim10\units{\%}$ of dc input power to ac output power including the losses on the way from the generator to the on-chip detector and the power lost in higher harmonics. The typical measured linewidth of a free-running semifluxon generator is $\sim1$--$10\units{MHz}$ at a typical operating frequency of about $200\units{GHz}$.  The frequency and linewidth are not sensitive to the deviations of $I_\mathrm{inj}$ from $I_\mathrm{inj}^\pi$, \ie, to the value of the phase discontinuity. We have also demonstrated the possibility to phase lock the oscillator with the reference generator frequency, which collapses the linewidth theoretically to zero having the spectral ratio of 95\%. Thus, this type of Josephson oscillator is comparable with those based on flux-flow and even has some advantages, such as smaller size and insensitivity to $I_\mathrm{inj}$ (\ie, $I_\mathrm{CL}$ in the case of flux-flow).

This work was supported by the RFBR and the Ministry of Education and
Science of the Russian Federation (agreement 8641).

\bibliography{SF,LJJ,LJJ-stacks,JJ}

\end{document}